\documentclass{ws-procs9x6}

\def\etal {{\it et al.}}

\begin{document}

\title{TESTING LORENTZ INVARIANCE BY COMPARING\\
LIGHT PROPAGATION IN VACUUM AND MATTER\\
}

\author{M.\ NAGEL,$^{1*}$ K.\ M\"{O}HLE,$^1$ K.\ D\"{O}RINGSHOFF,$^1$ S.\ HERRMANN,$^{1,2}$ A.\ SENGER,$^1$ E.V.\ KOVALCHUK,$^1$ and A.\ PETERS$^1$}

\address{$^1$Institut f\"{u}r Physik, Humboldt-Universit\"{a}t zu Berlin\\
Newtonstr. 15, 12489 Berlin, Germany\\
$^2$ ZARM, Universit\"{a}t Bremen\\
Am Falltrum 1, 28359 Bremen, Germany\\
$^*$E-mail: moritz.nagel@physik.hu-berlin.de}

\begin{abstract}
We present a Michelson-Morley type experiment for testing the isotropy of the speed of light in vacuum and matter. The experiment compares the resonance frequency of a monolithic optical sapphire resonator with the resonance frequency of an orthogonal evacuated optical cavity made of fused silica while the whole setup is rotated on an air bearing turntable once every 45 s. Preliminary results yield an upper limit for the anisotropy of the speed of light in matter (sapphire) of $\Delta c/c<4\times10^{-15}$, limited by the frequency stability of the sapphire resonator operated at room temperature. Work to increase the measurement sensitivity by more than one order of magnitude by cooling down the sapphire resonator to liquid helium temperatures (LHe) is currently under way.
\end{abstract}

\bodymatter

\section{Introduction}\label{sec:Intro}
Testing the isotropy of the speed of light serves as a sensitive test of special relativity and Lorentz invariance. The classic experiment to test the isotropy of the speed of light uses a Michelson interferometer and was first performed by A.A.\ Michelson in Berlin (1880) and Potsdam (1881), Germany. He was later joined by E.W.\ Morley to perform an improved experiment in Cleveland, Ohio (1887) \cite{MM87}. Modern Michelson-Morley type experiments use electromagnetic resonators to probe for Lorentz invariance violations and are generally based on comparing the resonance frequencies of two similar orthogonal resonators while either actively rotating the setup or relying solely on Earth's rotation \cite{Hall,Holger1,Wolf,Antonini,Sven,Stanwix,Eisele,Sven2}.

In case of a linear resonator, a relative frequency change is most generally described by $\delta\nu/\nu_0=\delta c/c_0-\delta L/L_0-\delta n/n_0$, where $\delta c/c_0$ denotes a relative change in the speed of light in vacuum along the optical path, $\delta L/L_0$ denotes a relative change in the length of the optical path, and $\delta n/n_0$ denotes a relative change in the index of refraction along the optical path. The magnitude of the different types of Lorentz violations, all three of which can occur in the case of spontaneous Lorentz symmetry breaking \cite{Holger2,Holger3,Holger4}, depend on the composition of the material the resonator is made of. Comparing the eigenfrequencies of two similar resonators made of the same material --- as has been done in all previous reported modern Michelson-Morley experiments --- makes it impossible to distinguish between the different types of Lorentz violation \cite{Ralf} and due to the substraction of the different types an overall Lorentz violating signal could even be suppressed or canceled. However, the material dependency makes it possible to distinguish between the different types of Lorentz violations by using dissimilar electromagnetic resonators.

In the past, we have combined results of an experiment performed in our laboratory in Berlin, Germany, consisting of linear optical resonators made of fused silica with mirrors made of BK7 with the results of an experiment performed by Stanwix \etal\ in Perth, Australia, consisting of whispering gallery microwave resonators made of sapphire in order to give separate bounds on the different types of Lorentz violations \cite{JoinedMM07}. We note that since the experiments have not been optimized for this kind of comparison and have not been synchronized timewise, not all in principle obtainable information of such a combined experiment could be resolved.

In our new setup, we compare the eigenfrequencies of two orthogonal evacuated linear optical fused-silica resonators with the eigenfrequency of a monolithic linear optical sapphire resonator in which the light is propagating in matter. We thus can directly compare light propagation in vacuum and matter and search for possible violations of the isotropy of the speed of light, which indicate Lorentz violations in matter. Moreover, this setup enables us in principle to give independent, simultaneous bounds on a broad range of Lorentz invariance violating coefficients in the Maxwell and Dirac sector in the framework of the Standard-Model Extension (SME) \cite{SME}.

In what follows we will give an overview of our new setup containing a monolithic optical sapphire resonator and two crossed evacuated optical fused-silica cavities followed by the first results obtained. We will also discuss improvements of this setup by cooling down the sapphire resonator to LHe temperatures. In the end we will give a brief outlook on current improvements of this setup and the next generation of modern Michelson-Morley type experiments.

\begin{figure}[t]
  \includegraphics[width=1\textwidth]{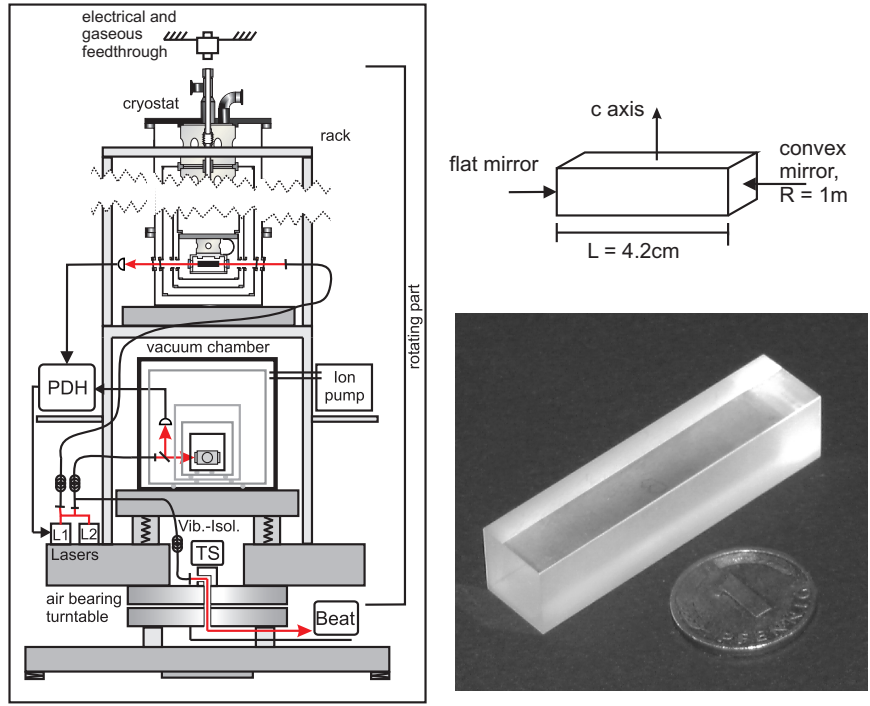}\\
  \caption{Right: schematic (top) and picture (bottom) of the monolithic sapphire resonator. Left: schematic of the new setup. The monolithic sapphire resonator is located in the cryostat at the upper level. The fused-silica resonators are located in the vacuum chamber at the lower level. PDH = Pound-Drever-Hall locking electronics. TS = tilt sensor.}\label{fig:Schema}
\end{figure}
\section{New setup}\label{sec:Extension}
We have realized a new type of a combined experiment in our laboratory in which we can compare the resonance frequency of a monolithic linear optical sapphire resonator with the resonance frequency of two evacuated linear optical cavities made of fused silica as used in our previous experiment \cite{Sven2} while actively rotating all resonators in a Michelson-Morley configuration on an air bearing turntable once every 45 s. Figure \ref{fig:Schema} shows a schematic of the sapphire monolithic resonator. Sapphire is a uniaxial crystal ($n_a=n_b\neq n_c$). The resonator was fabricated such that the crystal symmetry axis (c axis) lies perpendicular to the resonator axis. The eigenpolarizations are therefore linearly polarized, parallel to the crystal axis with extraordinary index $n_{eo}=n_c=1.7469$ at 1064 nm, and perpendicular to it with ordinary index $n_o=n_a=1.7546$. The modes are split in frequency due to the birefringence. The plane and convex (radius of curvature 1 m) end faces of the standing-wave resonator were coated with high-reflection (HR) coatings for a center wavelength of 1064 nm. The base material was a high-purity sapphire single-crystal rod (HEMEX ULTRA, Crystal Systems, Salem, MA).

The monolithic sapphire resonator features a finesse of about $10\,000$, corresponding to a linewidth of 200 kHz. The round trip loss inside the resonator is on the order of 600 ppm, although the loss due to absorption should only be on the order of $\sim10$ ppm/cm as measured by calorimetry. This leads to the conclusion that most of the losses are caused by flaw coatings. The incoupling efficiency of the monolithic sapphire resonator is less than $0.3\%$. For thermal shielding the sapphire resonator is placed inside a cryostat which, however, we operated at room temperature and pumped to a pressure of $<1\times 10^{-5}\,\text{mbar}$. It offers optical free beam access through windows (see Fig.\ \ref{fig:Schema}).

A Nd:YAG laser at 1064 nm is frequency stabilized to one of the TEM$_{00}$ modes of the resonator using a modified Pound-Drever-Hall method in transmission (modulation index of 3.75 and demodulating at third harmonic of the modulation frequency of 444 kHz). The incoupling light is first sent through a polarization maintaining optical fiber for mode cleaning and 20 mW of light power are impinging on the resonator's front side, resulting in $\sim20$ nW of transmitted light which is sensed for the Pound-Drever-Hall lock using an avalanche photodiode. Fractions of the laser beams of the laser stabilized to the sapphire resonator and the laser stabilized to one of the fused-silica resonators are overlapped on a fast photodiode to generate a beat note at the difference frequency $\Delta\nu=\left|\nu_{Sph}-\nu_{FS}\right|$.

The long term frequency drift of the beat note is normally $\sim2\,\text{kHz/s}$ in marked contrast to the drift of $<0.01\,\text{Hz/s}$ of the beat note between the two lasers stabilized to the fused-silica cavities \cite{Sven2}. This is due to the fact that there is no common mode rejection as in the case of the two crossed fused-silica cavities fabricated into one monolithic block. Additionally, the eigenfrequencies of the monolithic sapphire resonator feature a relatively high fractional temperature dependence at room temperature of $\Delta\nu/\nu_0 = -2.1\times 10^{-5} \Delta T/K$ resulting from its coefficient of thermal expansion (CTE) and the temperature dependance of the index of refraction at a wavelength of $\lambda=1064 \text{nm}$ of sapphire in contrast to $\Delta\nu/\nu_0 = -6\times 10^{-7} \Delta T/K$ for the fused-silica cavities resulting only from its CTE.
\begin{figure}[t]
  \includegraphics[width=1\textwidth]{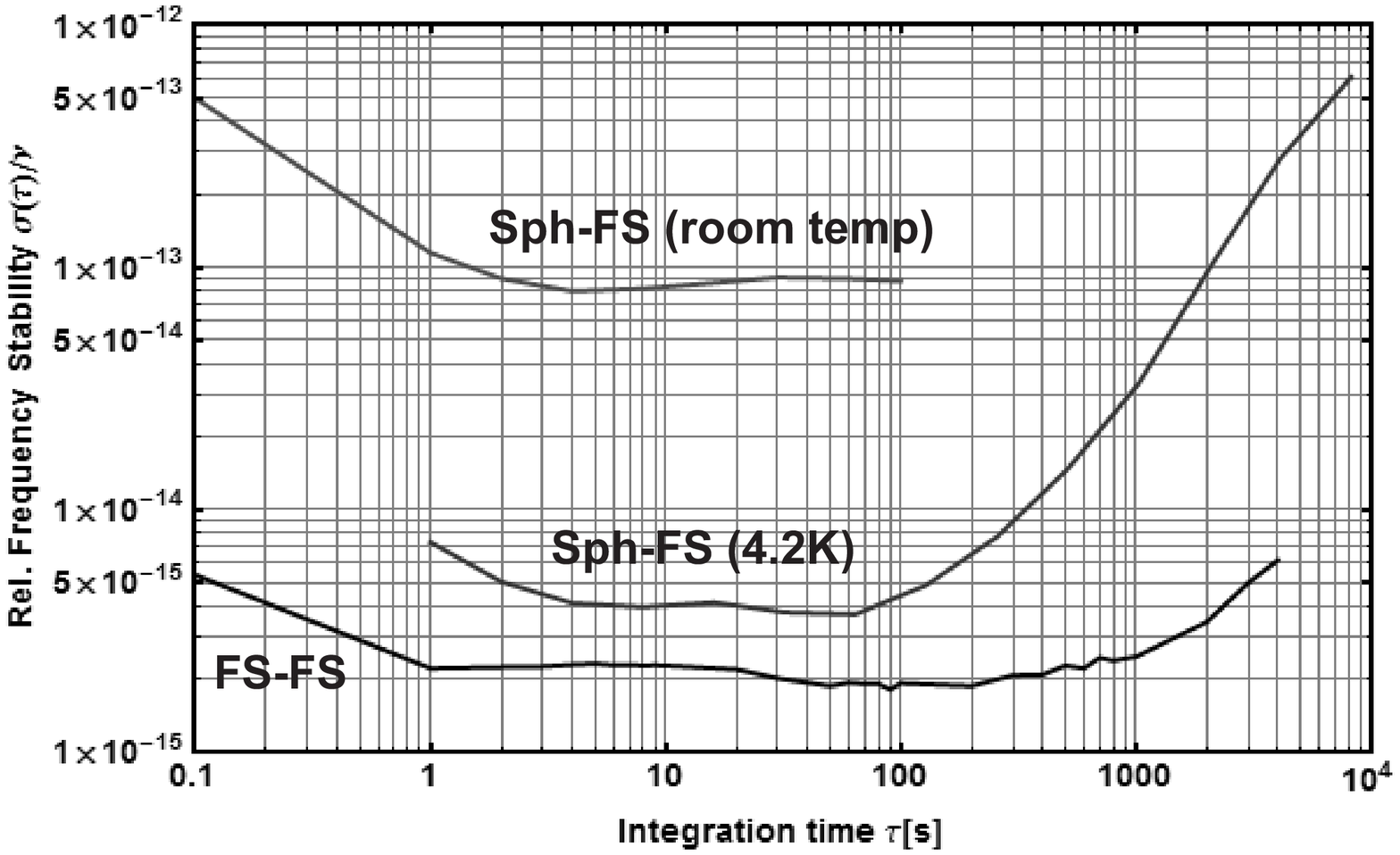}\\
  \caption{Relative frequency stability derived from the beat between the stabilized lasers (Sph = laser stabilized to the monolithic sapphire resonator, FS = laser stabilized to one of the fused-silica cavities).}\label{fig:AlaVar}
\end{figure}

The frequency stability of the beat note is characterized by the Allan deviation after removing the irrelevant long term drift by using the residuals of a linear plus quadratic fit (see Fig.\ \ref{fig:AlaVar}). There seems to be a flicker floor of $\sim8\times10^{-14}$ normalized to the laser frequency of 282 THz between 4 s and 100 s. We assume thermal noise within the monolithic sapphire to be the origin of this flicker floor, although this can not be affirmed by theory, as there is no complete theory on the thermal noise level of a monolithic optical resonator so far. However, we also cooled the monolithic sapphire resonator down to 4.2 K. The improvement in frequency stability at cryogenic temperatures make the assumption of thermal noise limiting effects reasonable (see Sec.\ \ref{sec:cryo}).

The vacuum chamber (cryostat) containing the monolithic sapphire resonator was placed on a breadboard containing all necessary optics. The breadboard itself was mounted to the rotating part of the existing setup above the vacuum chamber containing the crossed fused-silica resonators (see Fig.\ \ref{fig:Schema}) and thus represents a second new level within the setup. The sapphire resonator axis is orientated parallel to one of the fused silica's resonator axis and thus orthogonal to the resonator axis of the other fused-silica cavity. Except for these modifications are there no further changes of the existing setup and all measures implemented in our previous experiment \cite{Sven2} to reduce systematics connected with active rotation also apply for the monolithic sapphire resonator.

Ten days of comparison of the resonance frequency of the monolithic sapphire resonator operated at room temperature with the one of the orthogonal orientated fused-silica cavity while actively rotating at a chosen period of 45 s were performed in March 2009. This corresponds to more than $19\,000$ turntable rotations. Preparations for rotating measurements with the monolithic sapphire resonator cooled down to 4.2 K are currently under way (see Sec.\ \ref{sec:cryo}).

\section{First results}\label{sec:Results}
\begin{figure}[t]
  \includegraphics[width=1\textwidth]{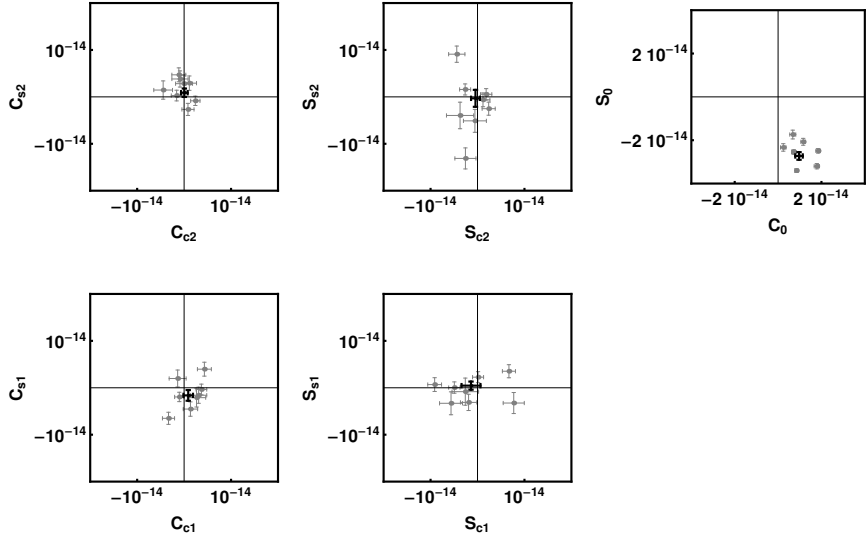}\\
  \caption{Modulation amplitudes (gray) and their mean values (black) as expected for an anisotropy of the speed of light fixed within a sidereal frame. Nomenclature as in our previous experiment \cite{Sven2}. Amplitudes $C_0$ and $S_0$ are most prone to constant systematic effects (note the different scale). The mean values and standard errors (one sigma) are $S_0 = -27.2\pm1.8$, $C_0 = 9.7\pm1.7$, $C_{s1} = -1.6\pm1.1$, $S_{s1} = 0.5\pm0.9$, $C_{c1} = 0.9\pm1.0$, $S_{c1} = -1.4\pm2.0$, $C_{s2} = 0.9\pm0.9$, $S_{s2} = -0.3\pm1.8$, $C_{c2} = 0.1\pm0.7$, $S_{c2} = -0.4\pm1.0$ (all values $\times 10^{-15}$).}\label{fig:Results}
\end{figure}

The analysis of the beat note with respect to anisotropy signals characterizing Lorentz invariance violations follows the same procedure as in our previous experiment \cite{Sven2}. No significant anisotropy signal was found fixed to a sidereal frame (see Fig.\ \ref{fig:Results}). Using the obtained sidereal modulation amplitudes we can conclude an upper limit for the anisotropy of the relative difference of the speed of light in vacuum and matter (sapphire) of $\Delta c/c = (3.8\pm2.4) \times10^{-15}$ (one standard deviation). A detailed analysis within the framework of the SME has not been done, since the dependence of the index of refraction of sapphire in the optical region on Lorentz violating coefficients of the photonic and fermionic sector has not been completely worked out yet. However, M\"{u}ller \cite{Holger4} has already described a recipe for deriving this dependency.

Further measurements with the monolithic sapphire resonator operated at room temperature have not been performed due to measurable residual systematic effects on the order of $\sim4\times10^{-14}$ acting on its eigenfrequencies at once and twice the rotation frequency. However, we note, that this anisotropy signal was fixed to the laboratory frame and averaged out when analyzing it with respect to a sidereal frame (see Fig.\ \ref{fig:Results}), which is the one of interest within the framework of the SME. The origin for anisotropy signals fixed to the laboratory are generally constant gradients (temperature, electromagnetic smog, gravity, etc.) that modulate the beat note at twice the turntable rotation rate.

The signals fixed to the laboratory in this case can be fully assigned to systematic effects influencing the eigenfrequencies of the monolithic sapphire resonator, since our previous experiments using the fused-silica cavities showed no significant signal \cite{Sven2}. The high round-trip loss rate and the high relative temperature dependency of the eigenfrequencies of the monolithic sapphire resonator can most likely explain the residual systematic effects. For example, the sensitivity of the monolithic sapphire resonator towards translational displacements of the incoupling beam --- which causes a change of the circulating light power inside the resonator and thus its temperature --- has been measured to be $\sim10\,\text{Hz}/\mu\text{m}$ at the timescale corresponding to twice the rotation rate. Since the resonator is free beam coupled with no active pointing stabilization, it is possible that the residual systematic effects fixed in the laboratory are caused by small deformations of the breadboard carrying the optics for the monolithic sapphire resonator system, while the setup is turning.

\section{Cryogenic operation}\label{sec:cryo}
Before continuing the measurements, we decided to minimize the residual systematic effects, which is most easily done by lowering the sensitivity to temperature changes of the eigenfrequencies of the sapphire resonator. This can be achieved by cooling down the sapphire resonator to cryogenic temperatures, e.g., 4.2 K. At these temperatures, the temperature sensitivity of the eigenfrequencies should decrease to $\sim4\times10^{-11}/$K as suggested by previous experiments using cryogenic sapphire resonators \cite{Seel} and thus pointing instabilities or laser light power fluctuations, etc.\ should have less impact on the overall measurement.

To enable operation at cryogenic temperatures, we changed the mounting of the monolithic sapphire resonator in the cryostat which first was just used as a vacuum chamber only (see Sec.\ \ref{sec:Extension}). At cryogenic temperatures an improvement of more than one order of magnitude in frequency stability for the eigenfrequencies of the monolithic sapphire resonator can be seen in the Allan deviation of the beat note (see Fig.\ \ref{fig:AlaVar}). Furthermore, a reduction in the beat note long term frequency drift down to 50 Hz/s was observed. A comparison with a third independent frequency reference, namely an actively temperature stabilized ULE cavity, reveled an  even lower long term frequency drift for the monolithic sapphire resonator of less than 0.1 Hz/s, where most of the drift can be assigned to some residual relaxation processes within the sapphire crystal due to the thermal cycling.

Preparations to perform rotating measurements with the monolithic sapphire resonator cooled down to LHe temperatures are currently under way. At cryogenic temperatures the measurement sensitivity will increase by more than one order of magnitude due to the better frequency stability of the sapphire resonator as compared to the room temperature case while the sensitivity to residual systematics will be lowered.

\section{Summary and outlook}\label{sec:Outlook}
We have set up a Michelson-Morley experiment which is capable of comparing light propagation in matter and vacuum. The derived limit for an anisotropy of the speed of light in vacuum in comparison with light propagation in matter (sapphire) is $\Delta c/c = (3.8\pm2.4)\times10^{-15}$ (preliminary). The measurement is currently limited by the frequency stability of the monolithic sapphire resonator operated at room temperature. Rotating measurements with the sapphire resonator cooled to LHe temperatures will soon be performed.

We are also planing the next generation of a modern Michelson-Morley experiment, in which we will overcome the momentarily limiting thermal noise limit \cite{Eisele,Sven2} by employing cryogenically cooled linear evacuated optical sapphire cavities. With these new cavities we should increase our measurement sensitivity for Lorentz invariance violations by about three orders of magnitude as compared to our previous experiment \cite{Sven2} and thus we will be able to probe the anisotropy of the speed of light in vacuum in the $10^{-20}$ regime. The improvement in measurement sensitivity will result from the potentially more than two orders of magnitude better frequency stability of the new sapphire cavities as compared to the fused-silica cavities used in our previous experiment \cite{Sven2}.

Moreover, we will combine our new experimental setup with cryogenic whispering gallery microwave resonators, which will be delivered and operated by the University of Western Australia \cite{Stanwix}, where light is propagating in matter and that feature a frequency stability in the mid $10^{-16}$ region. With the proposed co-rotating microwave and optical resonator setup we will be able to probe for possible Lorentz violations that are either frequency dependent (microwave vs.\ optical), rely on the electromagnetic mode (whispering gallery vs.\ linear), and/or depend on the medium the light is propagating in (matter vs.\ vacuum). We thus will also be able to set first limits on a broad range of Lorentz violating coefficients within the framework of the recently formulated extended SME \cite{Mewes2}.

\end{document}